\documentclass[preprint,showpacs,aps,prc,superscriptaddress]{revtex4-2}

\usepackage{graphicx}
\usepackage{amsmath,amssymb}
\usepackage[hidelinks]{hyperref}
\usepackage{multirow}
\usepackage{booktabs}
\usepackage{braket}
\usepackage[dvipsnames]{xcolor}
\usepackage{tikz}
\usetikzlibrary{calc}

\newcommand{\thetaz}{\Theta^{+}}

\newcommand{\DiagDeliveredPions}{8.64\times10^{11}}

\newcommand{\DiagEOneSigmaUL}{0.0293}

\newcommand{\DiagEOneFive}{0.0894}
\newcommand{\DiagResonanceSigmaUL}{0.0356}

\newcommand{\DiagResonanceFive}{0.109}

\newcommand{\DiagDensityBackground}{3.77\times10^{5}}
\newcommand{\DiagResonanceFWHM}{3.56}

\newcommand{\DiagEOneWindowMin}{0.844}
\newcommand{\DiagEOneWindowMax}{1.04}
\newcommand{\DiagResonanceWindowMin}{1.42}
\newcommand{\DiagResonanceWindowMax}{1.74}

\newcommand{\DiagFormationBeam}{2.586}

\newcommand{\DiagParityTemplateShiftPercent}{0.1}
\newcommand{\DiagParityKRatioMax}{50}
\newcommand{\DiagPosWidthLowMin}{0.0039}
\newcommand{\DiagPosWidthLowMax}{0.0634}
\newcommand{\DiagPosWidthFormMin}{0.00664}
\newcommand{\DiagPosWidthFormMax}{0.079}
\newcommand{\DiagNegWidthLowMin}{0.0617}
\newcommand{\DiagNegWidthLowMax}{0.383}
\newcommand{\DiagNegWidthFormMin}{0.045}
\newcommand{\DiagNegWidthFormMax}{1.47}
\newcommand{\DiagNegPSStaticLow}{0.196}
\newcommand{\DiagNegPSStaticForm}{0.333}
\newcommand{\DiagNegPSCovLow}{0.383}
\newcommand{\DiagNegPSCovForm}{1.47}

\begin{document}

\title{Sensitivity to \texorpdfstring{$\Theta^{+}$}{Theta+} production in the \texorpdfstring{$\pi^{-}p\to K^{-}X$}{pi- p to K- X} reaction at HIAF-HIRIBL}

\author{Yu~Xu}
\email{xuyu@gdlhz.ac.cn}
\affiliation{Institute of Modern Physics, Chinese Academy of Sciences, Lanzhou 730000, China}
\affiliation{Advanced Energy Science and Technology Guangdong Laboratory, Huizhou 516000, China}
\author{Ju-Jun~Xie}
\affiliation{Institute of Modern Physics, Chinese Academy of Sciences, Lanzhou 730000, China}
\affiliation{Advanced Energy Science and Technology Guangdong Laboratory, Huizhou 516000, China}
\affiliation{State Key Laboratory of Heavy Ion Science and Technology, Institute of Modern Physics, Chinese Academy of Sciences, Lanzhou 730000, China}
\affiliation{School of Nuclear Science and Technology, University of Chinese Academy of Sciences, Beijing 100049, China}
\affiliation{Southern Center for Nuclear-Science Theory (SCNT), Institute of Modern Physics, Chinese Academy of Sciences, Huizhou 516000, Guangdong, China}
\author{Xueheng~Zhang}
\email{zhangxh@impcas.ac.cn}
\affiliation{Institute of Modern Physics, Chinese Academy of Sciences, Lanzhou 730000, China}
\affiliation{Advanced Energy Science and Technology Guangdong Laboratory, Huizhou 516000, China}
\affiliation{State Key Laboratory of Heavy Ion Science and Technology, Institute of Modern Physics, Chinese Academy of Sciences, Lanzhou 730000, China}
\affiliation{School of Nuclear Science and Technology, University of Chinese Academy of Sciences, Beijing 100049, China}
\author{Liangwen~Chen}
\affiliation{Institute of Modern Physics, Chinese Academy of Sciences, Lanzhou 730000, China}
\affiliation{Advanced Energy Science and Technology Guangdong Laboratory, Huizhou 516000, China}
\affiliation{School of Nuclear Science and Technology, University of Chinese Academy of Sciences, Beijing 100049, China}
\author{Yuhong~Yu}
\affiliation{Institute of Modern Physics, Chinese Academy of Sciences, Lanzhou 730000, China}
\affiliation{Advanced Energy Science and Technology Guangdong Laboratory, Huizhou 516000, China}
\affiliation{School of Nuclear Science and Technology, University of Chinese Academy of Sciences, Beijing 100049, China}
\author{Li~Deng}
\affiliation{Institute of Modern Physics, Chinese Academy of Sciences, Lanzhou 730000, China}
\affiliation{Advanced Energy Science and Technology Guangdong Laboratory, Huizhou 516000, China}
\author{Jiajia~Zhai}
\affiliation{Institute of Modern Physics, Chinese Academy of Sciences, Lanzhou 730000, China}
\affiliation{Advanced Energy Science and Technology Guangdong Laboratory, Huizhou 516000, China}
\author{Pei~Yu}
\affiliation{Institute of Modern Physics, Chinese Academy of Sciences, Lanzhou 730000, China}
\affiliation{Advanced Energy Science and Technology Guangdong Laboratory, Huizhou 516000, China}
\author{He~Zhao}
\affiliation{Institute of Modern Physics, Chinese Academy of Sciences, Lanzhou 730000, China}
\affiliation{Advanced Energy Science and Technology Guangdong Laboratory, Huizhou 516000, China}
\affiliation{School of Nuclear Science and Technology, University of Chinese Academy of Sciences, Beijing 100049, China}
\author{Lina~Sheng}
\affiliation{Institute of Modern Physics, Chinese Academy of Sciences, Lanzhou 730000, China}
\affiliation{Advanced Energy Science and Technology Guangdong Laboratory, Huizhou 516000, China}
\affiliation{School of Nuclear Science and Technology, University of Chinese Academy of Sciences, Beijing 100049, China}
\author{Guodong~Shen}
\affiliation{Institute of Modern Physics, Chinese Academy of Sciences, Lanzhou 730000, China}
\affiliation{Advanced Energy Science and Technology Guangdong Laboratory, Huizhou 516000, China}
\affiliation{School of Nuclear Science and Technology, University of Chinese Academy of Sciences, Beijing 100049, China}

\author{Zhiyu~Sun}
\affiliation{Institute of Modern Physics, Chinese Academy of Sciences, Lanzhou 730000, China}
\affiliation{Advanced Energy Science and Technology Guangdong Laboratory, Huizhou 516000, China}
\affiliation{State Key Laboratory of Heavy Ion Science and Technology, Institute of Modern Physics, Chinese Academy of Sciences, Lanzhou 730000, China}
\affiliation{School of Nuclear Science and Technology, University of Chinese Academy of Sciences, Beijing 100049, China}

\date{\today}

\begin{abstract}
Pion-induced searches for the $\Theta^+$ can probe the proposed $N^*(2400)$ formation region beyond the E19 energies. We present sensitivity projections for a missing-mass search for a narrow $\Theta^+$ at $1.540$~GeV/$c^2$ in $\pi^-p\to K^-X$ at the High Intensity heavy-ion Accelerator Facility and its HIgh-rigidity Radioactive Ion Beam Line. With the E19 resolutions, efficiency, and continuum, separate 10-day exposures at pion momenta of $2.01$ and $\DiagFormationBeam$~GeV/$c$ give background-only median expected 90\% CL upper limits of $\DiagEOneSigmaUL$ and $\DiagResonanceSigmaUL~\mu$b/sr, respectively, on the differential cross section averaged over $2^\circ$--$15^\circ$ on the selected high-momentum branch; the higher momentum corresponds to the proposed $N^*(2400)$ formation energy. Median local $5\sigma$ discovery thresholds are $\DiagEOneFive$ and $\DiagResonanceFive~\mu$b/sr. Nonresonant nucleon-exchange models give width scales of $\DiagPosWidthLowMin$--$\DiagPosWidthFormMax$~MeV for $J^P=1/2^+$ and $\DiagNegWidthFormMin$--$\DiagNegWidthFormMax$~MeV for $J^P=1/2^-$. All positive-parity variants would test the DIANA width scale of $0.34\pm0.10$~MeV; negative-parity sensitivity depends on the variant.
\end{abstract}

\maketitle

\section{Introduction}
\label{sec:intro}

The $\thetaz$ is a candidate pentaquark state with strangeness $S=+1$ and minimal quark content $uudd\bar s$. Its predicted narrow width~\cite{Diakonov1997}, constrained to the few-MeV level or below by analyses of $K^+N$ and $K^+d$ data~\cite{Arndt2003,Sibirtsev2004}, and the first photoproduction signal reported by LEPS~\cite{Nakano2003} prompted both production and formation searches. Photoproduction signals were subsequently reported by SAPHIR~\cite{Barth2003} and by HERMES in quasi-real photoproduction~\cite{Airapetian2004}, and a further signal was reported in a higher-statistics LEPS measurement~\cite{Nakano2009}; a $K^0_sp$ structure was seen by ZEUS in deep-inelastic scattering~\cite{Chekanov2004}, but not confirmed in its later search with $358$~pb$^{-1}$~\cite{Abramowicz2016}; the CLAS measurements on hydrogen and deuterium found no significant peak~\cite{DeVita2006,McKinnon2006}, nor did the Belle and COSY-TOF searches~\cite{Mizuk2006,AbdelBary2007}. A Bayesian reanalysis of the CLAS data found no significant evidence for the state~\cite{Ireland2008}, although its statistical interpretation has been disputed~\cite{Cousins2008}, while the DIANA formation measurement gave a positive signal~\cite{Diana2014}. Diquark-based pictures of the state have been proposed~\cite{Jaffe2003,Karliner2003}, and its mass and parity have been addressed in lattice calculations~\cite{Sasaki2004}. The discovery of hidden-charm pentaquarks at LHCb revived interest in multiquark states~\cite{Aaij2015,Aaij2016,Aaij2019}, but the existence of the $S=+1$ $\Theta^+$ remains unestablished~\cite{Hicks2012,Amaryan2022,Amaryan2025}.

For a narrow-state search, pion-induced production on hydrogen offers the combination of high-intensity secondary pion beams and few-MeV missing-mass resolution. On a hydrogen target, the missing mass in $\pi^-p\to K^-X$ can be reconstructed from the four-momenta of the incident pion and outgoing kaon without detecting the $\Theta^+$ decay products. Using a free proton also avoids corrections for nuclear Fermi motion and intranuclear rescattering. J-PARC E19 measured this channel at $p_\pi=1.92$~GeV/$c$ in 2010 and $2.01$~GeV/$c$ in 2012~\cite{Moritsu2014,Shirotori2012}. The first setting was chosen to reexamine the weak enhancement reported by KEK E522 at the same beam momentum~\cite{Miwa2006}. The second used the maximum momentum of the K1.8 beamline to explore the larger production cross section anticipated by the reaction models~\cite{Oh2004,Hyodo2012}. The missing-mass resolutions (FWHM) were $1.72$~MeV at $1.92$~GeV/$c$ and $2.13$~MeV at $2.01$~GeV/$c$~\cite{Moritsu2014}. No significant narrow peak was observed at either setting. The 90\% confidence-level upper limits on $\Theta^+$ production, averaged over laboratory angles $2^\circ$--$15^\circ$, were at most $0.28~\mu$b/sr over $1510$--$1550$ and $1500$--$1560$~MeV/$c^2$ at $1.92$ and $2.01$~GeV/$c$, respectively~\cite{Moritsu2014}. These measurements provide the benchmark for pion-induced $\Theta^+$ production, but cover center-of-mass energies only up to about $2.16$~GeV.

The CLAS measurement of $\gamma p\to\pi^+K^-K^+n$ reported an excess near $2.4$~GeV/$c^2$ in the $K^-\Theta^+$ invariant-mass distribution for candidate $\Theta^+$ events~\cite{Kubarovsky2004}. Karliner and Lipkin interpreted this structure as a possible cryptoexotic $N^*(2400)$ with hidden strangeness~\cite{Karliner2004}. Formation of this state in $\pi^-p$ collisions has been considered~\cite{Azimov2004,Naruki2016}. The corresponding formation energy, $\sqrt{s}\simeq2.4$~GeV, lies beyond the range investigated by E19. Measurements of the $\Theta^+$ production excitation function in this region would test the proposed mechanism.

The High Intensity heavy-ion Accelerator Facility (HIAF) and its HIgh-rigidity Radioactive Ion Beam Line (HIRIBL) offer a potential means of studying pion-induced $\Theta^+$ production both in the E19 energy range and at higher energies~\cite{HIAF,Sheng2020,Wang2025}. This work quantifies the projected cross-section reach in the E19 energy region and near the proposed $N^*(2400)$ formation energy at $p_\pi\simeq2.586$~GeV/$c$. Expected upper limits and discovery thresholds follow from missing-mass spectra. Effective Lagrangian calculations~\cite{Oh2004,Hyodo2012} provide angular distributions and, within the nonresonant nucleon-exchange description, the model-dependent conversion from a production cross section to a $\Theta^+$ width. Section~\ref{sec:simulation} describes the facility, measurement configuration, and sensitivity calculation. The cross-section reach, its implications for the $\Theta^+$ width, and the test of $N^*(2400)$ formation are discussed in Sec.~\ref{sec:results}. The production amplitudes and kinematic transformations are collected in the Appendix.

\section{Measurement model and sensitivity calculation}
\label{sec:simulation}
\label{sec:sensitivity}

\subsection{HIAF and the HIRIBL beamline}
\label{sec:facility}
\label{sec:hfrs}

\begin{figure*}[t]
\centering
\includegraphics[width=\textwidth]{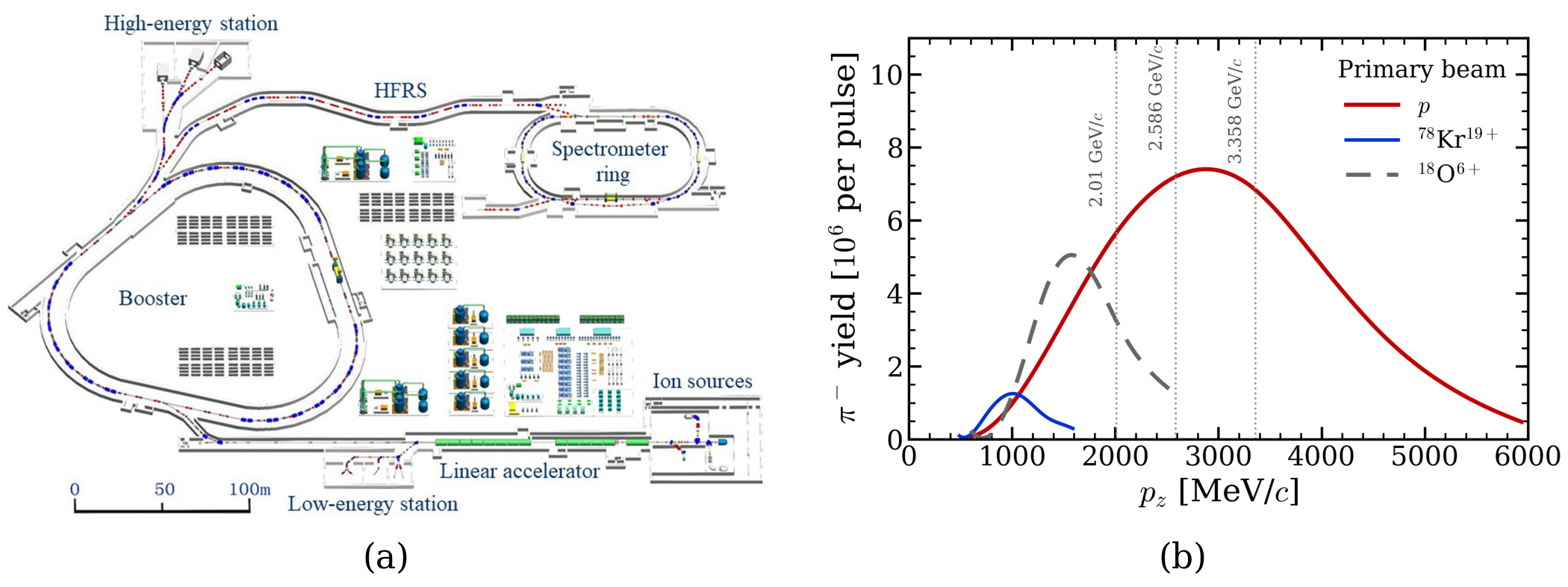}
\caption{(a) Layout of HIAF~\cite{HIAF}. (b) $\pi^-$ yield at the exit of HIRIBL for the proton, $^{78}$Kr$^{19+}$ and $^{18}$O$^{6+}$ beams.}
\label{fig:facility}
\end{figure*}

HIAF is a high-intensity heavy-ion accelerator facility in Huizhou, China. Construction was completed in late 2025, and commissioning began in 2026. The layout of the facility is shown in Fig.~\ref{fig:facility}(a). For the proposed measurement, the primary beam is extracted onto a production target. The secondary pions are momentum-selected by HIRIBL and transported to the experimental target. HIRIBL, formerly known as the High energy FRagment Separator (HFRS), is a two-stage in-flight separator with an optical length of about $192$~m and a maximum rigidity of $15$~T~m in Phase-I~\cite{HIRIBL2024}, with a design value of $25$~T~m~\cite{Sheng2020}. It comprises a pre-separator, which rejects the primary beam and unwanted reaction products, and a main separator, which provides further selection and transport. For singly charged particles, the Phase-I rigidity corresponds to a maximum momentum of about $4.5$~GeV/$c$; the design rigidity would extend this to $7.5$~GeV/$c$~\cite{HIAFmuon2024}. The relevant optical parameters are summarized in Table~\ref{tab:hfrs_params}.

We calculate pion production and subsequent transport through HIRIBL using the same primary-beam and beamline parameters as in the HIAF muon-source study of Ref.~\cite{HIAFmuon2024}. Figure~\ref{fig:facility}(b) shows the resulting $\pi^-$ yield at the beamline exit for the proton beam and for the $^{78}$Kr$^{19+}$ and $^{18}$O$^{6+}$ alternatives. With the proton beam the pion yield exceeds $5\times10^6$ per pulse between $2$ and $4$~GeV/$c$, with a maximum near $3$~GeV/$c$. These yields indicate that secondary pion beams with momenta above the E19 range could be available.

In this work the sensitivity is evaluated at two pion beam momenta, $2.01$ and $\DiagFormationBeam$~GeV/$c$. The first matches the E19 beam momentum, so the projection is made under the same beam conditions (Sec.~\ref{sec:results}). The second corresponds to $\sqrt{s}=2.4$~GeV, the formation energy of the proposed $N^*(2400)$, and probes $K^-\Theta^+$ production through that channel.

\begin{table}[t]
\centering
\caption{Optical parameters of the HIRIBL beamline~\cite{Sheng2020,HIRIBL2024}.}
\label{tab:hfrs_params}
\begin{ruledtabular}
\begin{tabular}{lc}
Parameter & Value \\
\hline
Optical length & $192$~m \\
Maximum rigidity & $15$~T~m (Phase-I); $25$~T~m (design) \\
Angular acceptance & $\pm30$~mrad (horizontal), $\pm25$~mrad (vertical) \\
Momentum acceptance & $\pm2\%$ \\
\end{tabular}
\end{ruledtabular}
\end{table}

\subsection{Measurement configuration and kinematics}
\label{sec:configuration}
\label{sec:kinematics}

For a stationary proton target,
\begin{equation}
s=m_\pi^2+M_N^2+2M_N\sqrt{p_\pi^2+m_\pi^2}.
\label{eq:fixed_target}
\end{equation}
Here $m_\pi$ and $M_N$ are the pion and proton masses. The two beam momenta, $2.01$ and $\DiagFormationBeam$~GeV/$c$, correspond to $\sqrt{s}=2.16$ and $2.40$~GeV, respectively.

We assume a conservative reference exposure at both momenta: $5\times10^6$ identified pions delivered to the liquid-hydrogen target per pulse, a 5-s pulse cycle, and 10 days of beam time, i.e.\ $\DiagDeliveredPions$ pions in total and a cycle-averaged target-delivered rate of $10^6$~s$^{-1}$.

We measure the incident pion momentum event by event. A dedicated downstream magnetic spectrometer measures the scattered kaon. We assume a uniform geometrical acceptance over $2^\circ\leq\theta\leq15^\circ$ and an azimuthal sector of $2.99$~rad, giving $0.100$~sr, with a 4.5-m kaon flight path and a target areal density of $0.86$~g/cm$^2$. We use kaon momentum windows of $\DiagEOneWindowMin$--$\DiagEOneWindowMax$ and $\DiagResonanceWindowMin$--$\DiagResonanceWindowMax$~GeV/$c$ at $2.01$ and $\DiagFormationBeam$~GeV/$c$, respectively, each set by the corresponding signal momentum envelope with a 5\% margin at each end. We adopt the E19 total detection efficiency of $15.1\%$~\cite{Moritsu2014}, replacing only its representative kaon-decay factor by the momentum-dependent survival $\exp[-Lm_K/(p_Kc\tau_K)]$ evaluated at the kaon momenta of the two settings.

The missing mass is reconstructed from the measured pion and kaon momenta and their laboratory opening angle, with $E_{\pi,K}=\sqrt{p_{\pi,K}^2+m_{\pi,K}^2}$ and $m_K$ the kaon mass:
\begin{equation}
M_X^2=(E_\pi+M_N-E_K)^2-p_\pi^2-p_K^2+2p_\pi p_K\cos\theta.
\label{eq:missing}
\end{equation}
The kaon momentum at a given laboratory angle follows from two-body energy--momentum conservation. At $2.01$~GeV/$c$ the two-body kinematics is double valued over the accepted angular range; the lower solution lies near $0.11$~GeV/$c$, so the E19 $0.8$--$1.2$~GeV/$c$ window retains the high-momentum branch, as do the windows used here. The angular transformation and the branch summation used in the acceptance integrals are given by Eqs.~(\ref{eq:boost}) and (\ref{eq:jacobian}).

\subsection{Missing-mass spectra and sensitivity estimation}
\label{sec:spectra}

The intrinsic $\Theta^+$ width is neglected relative to the detector resolution, and the true signal mass is fixed at $M_\Theta=1.540$~GeV/$c^2$. Independent Gaussian measurement errors are propagated from the E19 reference inputs $\Delta p_\pi/p_\pi=1.4\times10^{-3}$, $\Delta p_K/p_K=2.0\times10^{-3}$, and $\Delta\theta=5.7$~mrad, all in FWHM~\cite{Moritsu2014}. After conversion to standard deviations, the local mass resolution is
\begin{equation}
\sigma_M^2=
 \left(\frac{\partial M_X}{\partial p_\pi}\right)^2\sigma_{p_\pi}^2+
 \left(\frac{\partial M_X}{\partial p_K}\right)^2\sigma_{p_K}^2+
 \left(\frac{\partial M_X}{\partial\theta}\right)^2\sigma_\theta^2+
 \sigma_{\rm strag}^2.
\label{eq:resolution}
\end{equation}
Here $\sigma_{\rm strag}$ denotes the energy-loss straggling contribution; it and the mass-scale bias are set to zero. For the reference pseudovector (PV) angular template, integration over the accepted angular distribution, weighted by the selection and efficiency factors, gives a mixture of local Gaussian responses. Its FWHM is $\DiagResonanceFWHM$~MeV at $\sqrt{s}=2.4$~GeV, compared with $2.03$~MeV at the E19 momentum; the larger width at the higher beam momentum reduces the sensitivity to a narrow state.

The production observable is the laboratory mean differential cross section on the selected high-momentum branch,
\begin{equation}
\bar\sigma=\frac{1}{\Omega_D}\int_D d\Omega_{\rm lab}\,
 \frac{d\sigma}{d\Omega_{\rm lab}},\qquad
\Omega_D=2\pi(\cos2^\circ-\cos15^\circ),
\label{eq:angular_mean}
\end{equation}
where $D$ covers the complete azimuth in the stated polar-angle interval. This angular average excludes apparatus acceptance, which enters the signal-yield calculation below. The dimensionless angular shape $g(\Omega)=(d\sigma/d\Omega_{\rm lab})/\bar\sigma$, with unit angular average, $\Omega_D^{-1}\int_D g(\Omega)\,d\Omega_{\rm lab}=1$, is calculated with the pseudoscalar (PS) and PV nucleon-exchange amplitudes of Ref.~\cite{Hyodo2012}, using the $J^P=1/2^+$ amplitudes unless stated otherwise.

For luminosity $\mathcal L=N_\pi n_t$, where $n_t$ is the target proton areal density, the expected signal in reconstructed-mass bin $i$ is $s_i=\bar\sigma C_i$, with
\begin{equation}
C_i=\mathcal L\int d\Omega_{\rm lab}\,
 g(\Omega)A(\Omega,p_K)\epsilon_s(\Omega,p_K)P_i(M_\Theta,\Omega).
\label{eq:signal_counts}
\end{equation}
Here $A$ specifies the selected angular sector and momentum window, and $\epsilon_s$ is the total detection efficiency. With luminosity in cm$^{-2}$, cross sections are converted from $\mu$b to cm$^2$. The probability of entering bin $[m_i,m_{i+1}]$ is
\begin{equation}
P_i(M,\Omega)=\Phi\!\left(\frac{m_{i+1}-M}{\sigma_M(M,\Omega)}\right)
 -\Phi\!\left(\frac{m_i-M}{\sigma_M(M,\Omega)}\right),
\label{eq:bin_probabilities}
\end{equation}
where $\Phi$ is the standard-normal cumulative distribution; $\sigma_M(M,\Omega)$ is evaluated at the kaon momentum corresponding to the true mass $M$ and laboratory angle $\Omega$.

The E19 background was dominated by $\phi$, $\Lambda(1520)$, and nonresonant $\bar KKN$ production~\cite{Moritsu2014}. The corresponding spectra at higher energies are not measured for this apparatus, so we adopt a genuine-kaon production density on the selected high-momentum branch, $\rho_b=0.66~\mu$b/sr/MeV over $1.44$--$1.64$~GeV/$c^2$, flat in laboratory angle. The value is the window average of the published E19 spectrum at $2.01$~GeV/$c$ digitized here; the observed variation over the window is $0.60$--$0.72~\mu$b/sr/MeV, which we treat as a background-normalization uncertainty, while shape freedom is carried by the fitted background deformation. It gives $\DiagDensityBackground$ selected background events in the $1.500$--$1.560$~GeV/$c^2$ fit window at $\sqrt{s}=2.4$~GeV. The expected reconstructed background is
\begin{equation}
b_i=\mathcal L\int dM\,d\Omega_{\rm lab}\,
 \rho_b(M,\Omega)A(\Omega,p_K(M))\epsilon_b(M,\Omega)P_i(M,\Omega).
\label{eq:background_counts}
\end{equation}
Signal and genuine-kaon background share the same detection efficiency. The produced density is held fixed across beam settings, while the kaon kinematics, momentum selection, and mass smearing are recomputed.

The sensitivity is evaluated in 120 bins of $0.5$~MeV/$c^2$ over $1.500$--$1.560$~GeV/$c^2$. Median expected limits use background-only Asimov data~\cite{Cowan2011}. Let $N=\bar\sigma C_{\rm all}\geq0$ denote the accepted signal yield integrated over all reconstructed masses. The likelihood in the fitted interval is
\begin{equation}
\mathcal L_{\rm P}(N,\boldsymbol\eta)=
 \prod_i {\rm Pois}\!\left(n_i\,\middle|\,
 Nt_i+b_i[1+\eta_0+\eta_1x_i+\eta_2x_i^2]\right),
\label{eq:likelihood}
\end{equation}
where $t_i=C_i/C_{\rm all}$, $\tilde m_i=(m_i+m_{i+1})/2$, and $x_i=(\tilde m_i-1.540~{\rm GeV})/(0.030~{\rm GeV})$; the sum of $t_i$ over the fitted bins is the retained signal fraction. The three background parameters are profiled freely, subject to positive background means. The nominal continuum therefore determines the expected counting rate, while the fit permits a local second-order deformation of its shape.

The median expected upper limit at 90\% CL is obtained with the CL$_{s}$ (modified-frequentist) prescription of Refs.~\cite{Read2002,Junk1999} by solving
\begin{equation}
2\left[1-\Phi\!\left(\sqrt{q_{N,A}}\right)\right]=0.10,
\qquad \bar\sigma_{90}=N_{90}/C_{\rm all},
\label{eq:expected_limit}
\end{equation}
where $q_{N,A}$ is the profile-likelihood statistic evaluated with background-only Asimov data, defined as minus twice the logarithm of the profile-likelihood ratio, and $N_{90}$ is the value of $N$ that solves Eq.~(\ref{eq:expected_limit}). The median expected discovery threshold at fixed mass is evaluated separately with signal-plus-background Asimov data, $Z_A=\sqrt{q_{0,A}}$.

Width bounds require a production model to convert cross-section limits. We use the nucleon-exchange model of Ref.~\cite{Hyodo2012} for both spin-parity assignments $J^P=1/2^{+}$ and $1/2^{-}$ of the $\Theta^+$, each with four combinations of pseudoscalar or pseudovector $KN\Theta$ vertices and static or covariant form factors. The amplitudes appear in the Appendix; Sec.~\ref{sec:results} defines $k^{C,F}$ for the same laboratory interval and selected branch.

\section{Results and discussion}
\label{sec:results}

\subsection{Projected cross-section sensitivity and model-dependent width constraints}
\label{sec:e19}
\label{sec:background_dependence}
\label{sec:validation_requirements}

\begin{figure}[t]
\centering
\includegraphics[width=0.85\columnwidth]{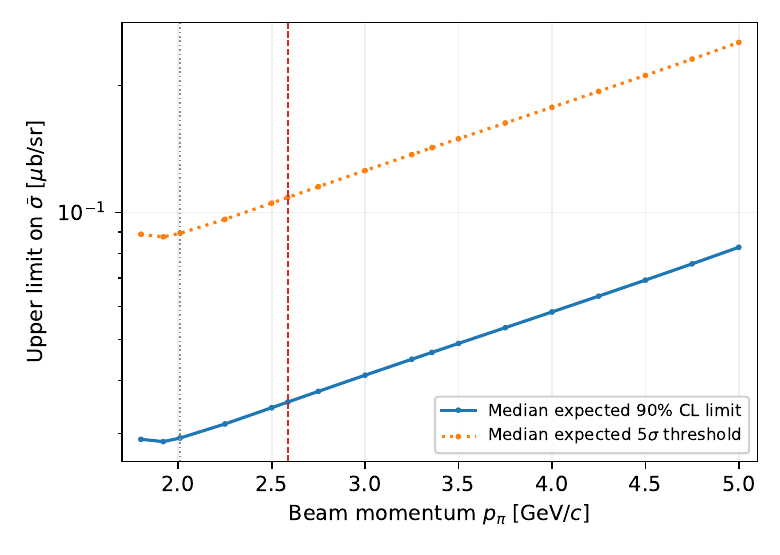}
\caption{Median expected $90\%$ CL upper limit on the angle-averaged production cross section and $5\sigma$ discovery threshold at fixed mass versus beam momentum, for a separate 10-day exposure at each momentum. The vertical lines mark the two beam settings of this work.}
\label{fig:sensitivity_comparison}
\end{figure}

\begin{figure}[t]
\centering
\includegraphics[width=\textwidth]{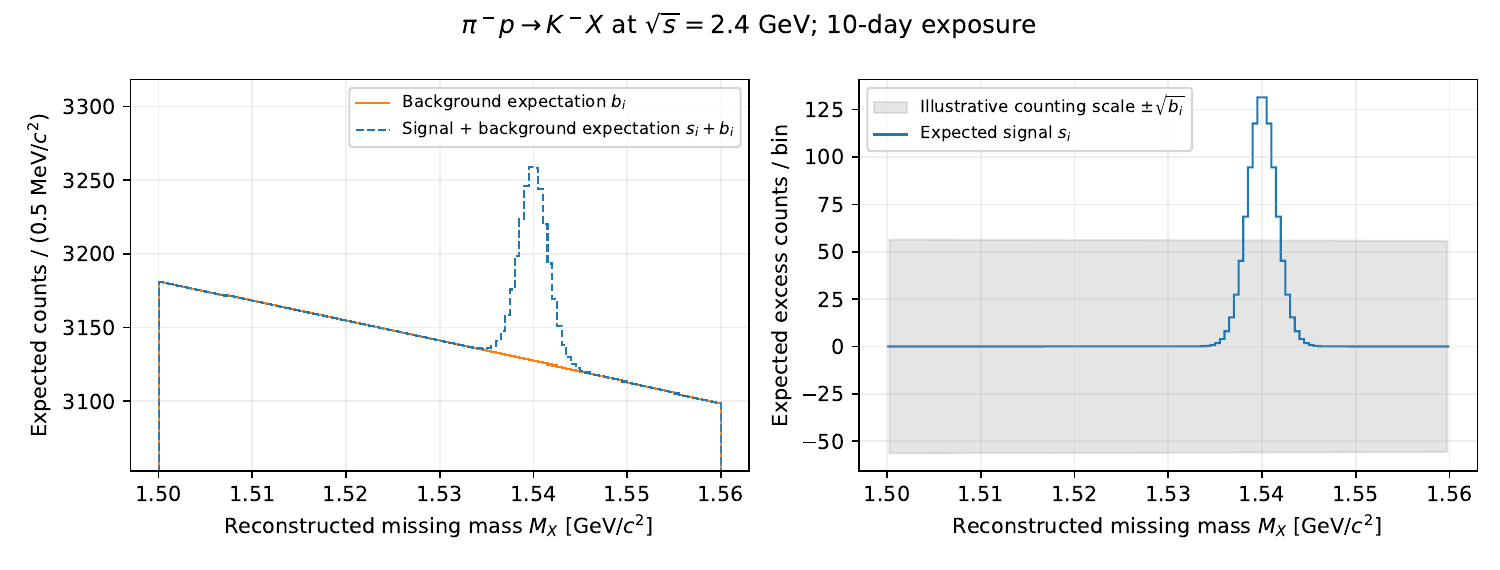}
\caption{Expected missing-mass spectrum at $\sqrt{s}=2.4$~GeV for the 10-day exposure, for a signal with $\bar\sigma=\DiagResonanceFive~\mu$b/sr at $M_\Theta=1.540$~GeV/$c^2$. Left: background $b_i$ and $b_i+s_i$. Right: the expected signal $s_i$, with the band showing $\pm\sqrt{b_i}$ counting fluctuations.}
\label{fig:expected_spectrum}
\end{figure}

For a 10-day exposure at $p_\pi=2.01$ and $2.586$~GeV/$c$, Table~\ref{tab:reach_benchmarks} gives background-only median expected 90\% CL upper limits of $\DiagEOneSigmaUL$ and $\DiagResonanceSigmaUL~\mu$b/sr on the angle-averaged differential production cross section, respectively. The corresponding cross sections yielding a median local significance of $5\sigma$ at the assumed mass are $\DiagEOneFive$ and $\DiagResonanceFive~\mu$b/sr. Figure~\ref{fig:sensitivity_comparison} shows the projected cross-section upper limits and $5\sigma$ discovery thresholds versus beam momentum; for fixed flux, efficiency, and continuum background density, the poorer missing-mass resolution at the higher momentum reduces sensitivity. Figure~\ref{fig:expected_spectrum} shows the corresponding expected missing-mass spectrum at $\sqrt{s}=2.4$~GeV for a signal with $\bar\sigma=\DiagResonanceFive~\mu$b/sr.

\begin{table}[t]
\centering
\caption{Projected cross-section sensitivity for separate 10-day exposures with the PV angular template.}
\label{tab:reach_benchmarks}
\small
\begin{tabular}{lrrr}
\hline
$p_\pi$ [GeV/$c$] & $\sqrt{s}$ [GeV] & 90\% CL [$\mu$b/sr] & $5\sigma$ [$\mu$b/sr] \\
\hline
2.01 & 2.16 & 0.0293 & 0.0894 \\
2.586 & 2.4 & 0.0356 & 0.109 \\
\hline
\end{tabular}

\end{table}

\label{sec:discussion}
\label{sec:beam_choice}

The angular-averaged production cross section is proportional to the $\Theta^+$ width within the model introduced in Sec.~\ref{sec:simulation},
\begin{equation}
\bar\sigma=k^{C,F}(p_\pi,M_\Theta)\Gamma_\Theta,
\label{eq:prop}
\end{equation}
where $C$ denotes the PS or PV vertex and $F$ the static or covariant form factor. The coefficient $k^{C,F}=\bar\sigma^{C,F}/\Gamma_\Theta$, in $(\mu{\rm b}/{\rm sr})/{\rm MeV}$, is evaluated over the selected branch and geometric angular interval of Eq.~(\ref{eq:angular_mean}), without detector efficiencies. Table~\ref{tab:k_coefficients} lists $k^{C,F}$ and the resulting 90\% CL width upper limits $\Gamma_{\Theta,90}=\bar\sigma_{90}/k^{C,F}$ at both beam momenta.

The conversion also depends on the spin parity assumed for the $\Theta^+$, through both the $\Theta KN$ vertex and the partial-wave relation that fixes $g_{KN\Theta}$ at a given width. We evaluate $J^P=1/2^+$ with the $p$-wave relation of Eq.~(\ref{eq:width_coupling}) and $J^P=1/2^-$ with the $s$-wave relation of Eq.~(\ref{eq:width_coupling_negative}); the corresponding amplitudes are given in the Appendix. Table~\ref{tab:k_coefficients_negative} repeats the four variants for $J^P=1/2^-$.

\begin{table}[t]
\centering
\caption{Born-model coefficients $k$ in $(\mu{\rm b}/{\rm sr})/{\rm MeV}$, evaluated for $J^P=1/2^+$ at $M_\Theta=1.540$~GeV/$c^2$ on the selected high-momentum branch, together with the 90\% CL width upper limits $\Gamma_{\Theta,90}=\bar\sigma_{90}/k$ in MeV for the 10-day exposures at both momenta.}
\label{tab:k_coefficients}
\label{tab:width_scales}
\small
\begin{tabular}{l|cc|cc}
\hline
 & \multicolumn{2}{c}{$p_\pi=2.01$~GeV/$c$} & \multicolumn{2}{c}{$p_\pi=2.586$~GeV/$c$} \\
\cline{2-3}\cline{4-5}
Variant & $k^{C,F}$ & $\Gamma_{\Theta,90}$ & $k^{C,F}$ & $\Gamma_{\Theta,90}$ \\
\hline
PS+$F_s$ & 7.51 & 0.0039 & 5.37 & 0.00664 \\
PS+$F_c$ & 3.85 & 0.00762 & 1.22 & 0.0292 \\
PV+$F_s$ & 0.901 & 0.0325 & 1.98 & 0.0179 \\
PV+$F_c$ & 0.461 & 0.0634 & 0.45 & 0.079 \\
\hline
\end{tabular}

\end{table}

\begin{table}[t]
\centering
\caption{Born-model coefficients $k$ in $(\mu{\rm b}/{\rm sr})/{\rm MeV}$, evaluated for $J^P=1/2^-$ at $M_\Theta=1.540$~GeV/$c^2$ on the selected high-momentum branch, together with the 90\% CL width upper limits $\Gamma_{\Theta,90}=\bar\sigma_{90}/k$ in MeV for the 10-day exposures at both momenta.}
\label{tab:k_coefficients_negative}
\small
\begin{tabular}{l|cc|cc}
\hline
 & \multicolumn{2}{c}{$p_\pi=2.01$~GeV/$c$} & \multicolumn{2}{c}{$p_\pi=2.586$~GeV/$c$} \\
\cline{2-3}\cline{4-5}
Variant & $k^{C,F}$ & $\Gamma_{\Theta,90}$ & $k^{C,F}$ & $\Gamma_{\Theta,90}$ \\
\hline
PS+$F_s$ & 0.149 & 0.196 & 0.107 & 0.333 \\
PS+$F_c$ & 0.0765 & 0.383 & 0.0243 & 1.47 \\
PV+$F_s$ & 0.475 & 0.0617 & 0.791 & 0.045 \\
PV+$F_c$ & 0.243 & 0.12 & 0.18 & 0.198 \\
\hline
\end{tabular}

\end{table}

For $J^P=1/2^+$, the four variants give width limits that differ by more than an order of magnitude, $\Gamma_{\Theta,90}=\DiagPosWidthLowMin$--$\DiagPosWidthLowMax$~MeV at $2.01$~GeV/$c$ and $\DiagPosWidthFormMin$--$\DiagPosWidthFormMax$~MeV at the formation energy.

For $J^P=1/2^-$ the projected limits are much weaker, $\Gamma_{\Theta,90}=\DiagNegWidthLowMin$--$\DiagNegWidthLowMax$~MeV at $2.01$~GeV/$c$ and $\DiagNegWidthFormMin$--$\DiagNegWidthFormMax$~MeV at the formation energy. At fixed $\Gamma_\Theta$ the $s$-wave relation of Eq.~(\ref{eq:width_coupling_negative}) gives $g_{KN\Theta}^2$ smaller by the factor $(E_N^d-M_N)/(E_N^d+M_N)\approx1/50$, so $k^{C,F}$ falls by about a factor of $\DiagParityKRatioMax$ for the PS vertices; the smaller mass difference $M_\Theta-M_N$ in the PV vertex compensates most of this and leaves a factor of $1.9$--$2.5$. The parity assignment therefore changes the width reach far more than the choice of beam momentum does. The cross-section limits themselves are insensitive to this choice: the angular templates of the two parities change $\bar\sigma_{90}$ by at most about $\DiagParityTemplateShiftPercent\%$.

\begin{figure}[t]
\centering
\includegraphics[width=\textwidth]{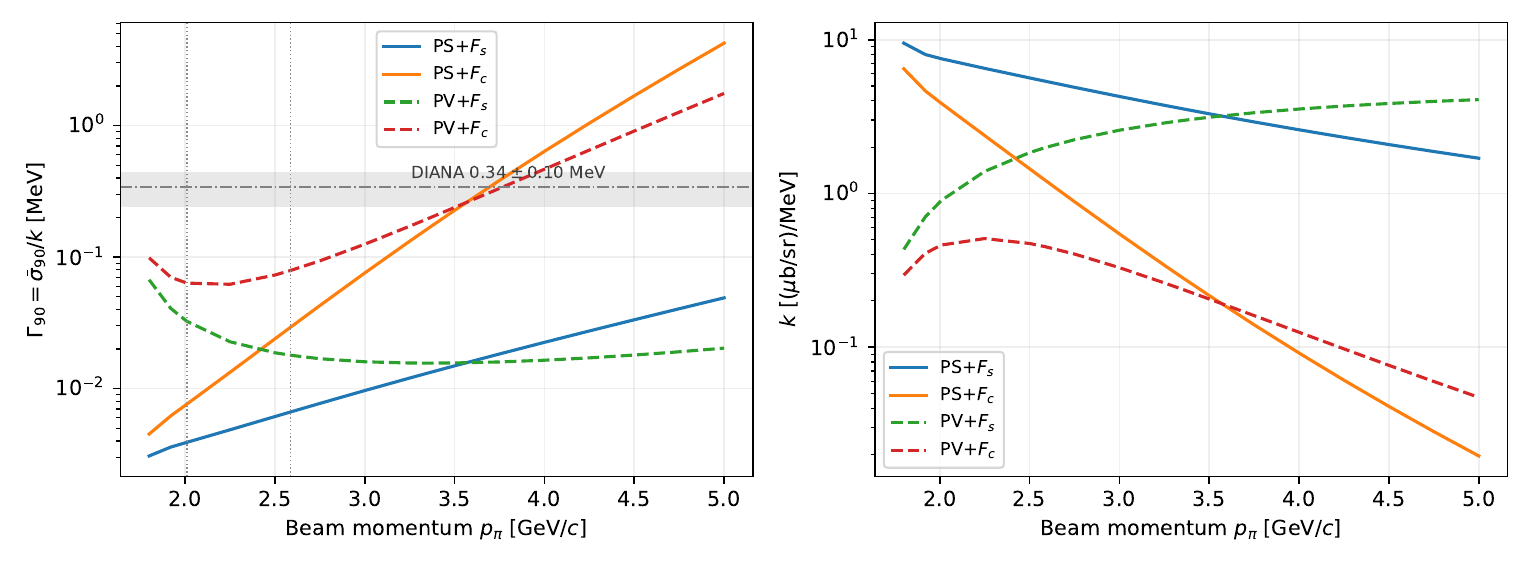}
\caption{Median expected $90\%$ CL width upper limit $\Gamma_{90}=\bar\sigma_{90}/k$ (left) and Born coefficient $k$ (right) versus beam momentum for $J^P=1/2^+$, for the four Born variants at fixed $10$-day exposure, flux, efficiency, and continuum density. The dotted vertical lines mark the two settings of this work. The dash-dotted line and shaded band show the width $0.34\pm0.10$~MeV reported by the DIANA Collaboration and its quoted uncertainty; they are not a $90\%$ allowed band.}
\label{fig:k_vs_p}
\end{figure}

\begin{figure}[t]
\centering
\includegraphics[width=\textwidth]{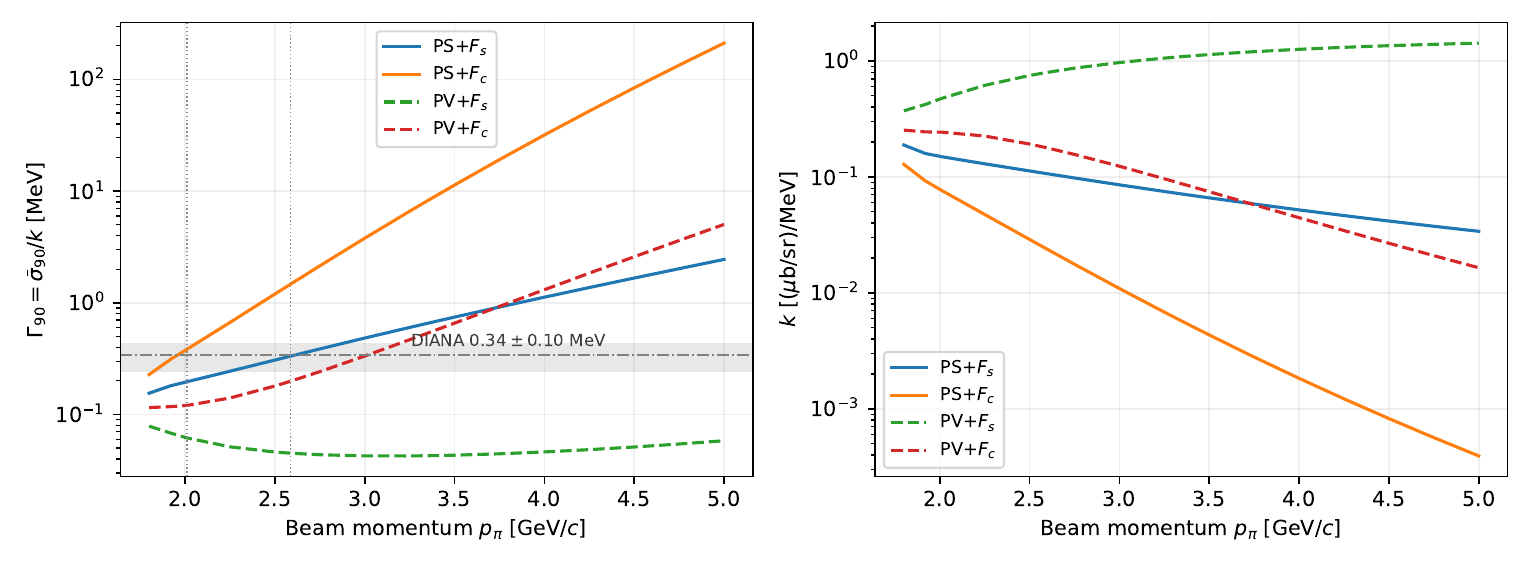}
\caption{Same as Fig.~\ref{fig:k_vs_p} for $J^P=1/2^-$. The projected width limits are larger because at a fixed width the $s$-wave relation of Eq.~(\ref{eq:width_coupling_negative}) gives a smaller coupling than the $p$-wave one.}
\label{fig:k_vs_p_negative}
\end{figure}

Figure~\ref{fig:k_vs_p} shows the beam-momentum dependence of $\Gamma_{90}=\bar\sigma_{90}/k$ (left) and of the Born coefficient $k$ (right) for $J^P=1/2^+$, and Fig.~\ref{fig:k_vs_p_negative} the same quantities for $J^P=1/2^-$. At both beam momenta, all four $J^P=1/2^+$ variants give background-only median expected 90\% CL width limits below $0.08$~MeV, so the proposed measurement would be sensitive to the width $0.34\pm0.10$~MeV reported by the DIANA Collaboration from $K^+$Xe data~\cite{Diana2014}. For $J^P=1/2^-$ the corresponding conversion scales are larger: PS+$F_s$ gives $\DiagNegPSStaticLow$ and $\DiagNegPSStaticForm$~MeV at $2.01$~GeV/$c$ and the formation energy, PS+$F_c$ gives $\DiagNegPSCovLow$ and $\DiagNegPSCovForm$~MeV, and both PV variants stay below the DIANA central value at both momenta. The $\DiagNegPSCovForm$~MeV entry is comparable to the $\DiagResonanceFWHM$~MeV mass resolution, so a finite-width signal response would be required to read it as a width limit. The quoted DIANA uncertainty does not include the systematic uncertainties of that extraction, and the comparison is conditional on its assumptions and on the Born model used here.

\subsection{Implications for \texorpdfstring{$N^*(2400)$}{N*(2400)} formation}
\label{sec:resonance}

The CLAS measurement of $\gamma p\to\pi^+K^-K^+n$ reported an excess near $2.4$~GeV/$c^2$ in the $K^-\Theta^+$ invariant-mass distribution for candidate $\Theta^+$ events~\cite{Kubarovsky2004}. Karliner and Lipkin interpreted this structure as a possible cryptoexotic $N^*(2400)$ with hidden strangeness~\cite{Karliner2004}. The pion-exchange interpretation of the photoproduction process motivated a direct search through $\pi^-p\to N^*(2400)\to K^-\Theta^+$ with a real pion beam~\cite{Azimov2004}. For $M_R=2.400$~GeV/$c^2$, Eq.~(\ref{eq:fixed_target}) gives $p_\pi=\DiagFormationBeam$~GeV/$c$, extending the search beyond the E19 energy range to the proposed formation region. For a narrow $\Theta^+$, Table~\ref{tab:reach_benchmarks} quantifies the median expected 90\% CL upper limit and the median local $5\sigma$ discovery threshold at the assumed mass for the angle-averaged production cross section, using the adopted PV angular template and detector response for a 10-day exposure. The resonant yield is governed by the $\pi NN^*$ and $K\Theta N^*$ couplings, which are not fixed by $\Gamma_\Theta$; consequently, the Born coefficients $k^{C,F}$ and the width conversion of Eq.~(\ref{eq:prop}) do not apply to this mechanism.

This sensitivity provides a quantitative benchmark for a search at the proposed formation energy: a null result would constrain $K^-\Theta^+$ production at this setting, and a narrow missing-mass signal would be consistent with the proposed decay channel. Establishing the intermediate $N^*(2400)$ requires measurements of the production excitation function across the formation region. Measurements on both sides of $\sqrt{s}\simeq2.4$~GeV, analyzed with coherent resonant and nonresonant amplitudes, would test whether the energy dependence supports a resonance contribution. Such a scan would connect the search for the $\Theta^+$ to the study of its production mechanism, complementing the proposed $K_Lp\to K^+n$ formation search at JLab~\cite{Amaryan2024KLF} and searches at the upgraded LEPS2 facility~\cite{Nakano2025LEPS2}.

\section{Summary}
\label{sec:conclusions}

The projected sensitivity of a $\pi^-p\to K^-X$ missing-mass search at HIAF-HIRIBL has been evaluated for separate 10-day exposures at $p_\pi=2.01$ and $\DiagFormationBeam$~GeV/$c$, under the adopted experimental assumptions. For a narrow $\Theta^+$ at $M_\Theta=1.540$~GeV/$c^2$ with the PV angular template, the background-only median expected 90\% CL upper limits on the production cross section averaged over the selected laboratory angular interval are $\DiagEOneSigmaUL$ and $\DiagResonanceSigmaUL~\mu$b/sr, respectively. The corresponding median expected $5\sigma$ discovery thresholds at fixed mass are $\DiagEOneFive$ and $\DiagResonanceFive~\mu$b/sr. At fixed flux, efficiency, and continuum density, the poorer missing-mass resolution at the higher beam momentum reduces the cross-section sensitivity.

The projected width scales $\Gamma_{\Theta,90}=\bar\sigma_{90}/k$ span $\DiagPosWidthLowMin$--$\DiagPosWidthFormMax$~MeV for $J^P=1/2^+$ and $\DiagNegWidthFormMin$--$\DiagNegWidthFormMax$~MeV for $J^P=1/2^-$, across the two beam settings and four model variants. Within the nucleon-exchange model, the median expected 90\% CL width limits for all four positive-parity variants and both negative-parity PV variants lie below the central value of the width $0.34\pm0.10$~MeV reported by the DIANA Collaboration, indicating sensitivity to this width scale at both beam settings. For negative parity, the PS+$F_s$ conversion scale lies below that central value at $2.01$~GeV/$c$ and is comparable to it at the formation energy, whereas the PS+$F_c$ scale remains above it at both settings. These projections would allow a model-dependent test of the DIANA width scale. Interpreting the largest conversion scale as a width limit requires a finite-width signal response. At $\sqrt{s}=2.4$~GeV, the projected cross-section sensitivity also provides a benchmark for $K^-\Theta^+$ production at the proposed $N^*(2400)$ formation energy, beyond the E19 energy range. The nucleon-exchange width conversion does not apply to this resonant mechanism, and establishing the intermediate state requires the production excitation function across the formation region.

\begin{acknowledgments}
The authors acknowledge discussions on HIAF beamline parameters with members of the HIRIBL collaboration. This work was financially supported by the Guangdong Basic and Applied Basic Research Foundation, China (Grant No.~2025A1515140112, 2023B1515120067), the National Natural Science Foundation of China (Grants Nos.~12105327, 12405337, 12405402, 12005221), the Guangdong Basic and Applied Basic Research Foundation, China (Grant No.~2023B1515120067), and the State Key Laboratory of Nuclear Physics and Technology, Peking University (Grant No.~NPT2025KFY07). Computing resources were provided by the supercomputing system in the Dongjiang Yuan Intelligent Computing Center.
\end{acknowledgments}

\appendix*
\makeatletter
\def\theequation@prefix{A}%
\makeatother
\section{Production amplitudes and kinematic transformations}
\label{app:born}

\subsection{Positive-parity nucleon-exchange amplitude}

For $J^P=1/2^+$, we use the $s$-channel nucleon Born amplitudes of Ref.~\cite{Hyodo2012}. In this subsection, $p_\pi$, $p_p$, and $p_K$ denote four-momenta and $P_s=p_\pi+p_p$; the spinors are normalized as $\bar uu=2M$, where $M$ is the corresponding baryon mass. Suppressing overall phases, the amplitudes are
\begin{align}
\mathcal M_0^{\rm PS}
 &=\frac{\sqrt2g_{\pi NN}g_{KN\Theta}}{s-M_N^2}
 \bar u_\Theta\gamma_5(\not P_s+M_N)\gamma_5u_p,\nonumber\\
\mathcal M_0^{\rm PV}
 &=\frac{\sqrt2g_{KN\Theta}g_A}
 {2f_\pi(M_\Theta+M_N)(s-M_N^2)}
 \bar u_\Theta\not p_K\gamma_5(\not P_s+M_N)\not p_\pi\gamma_5u_p.
\label{eq:born_amplitude}
\end{align}
We take $g_{\pi NN}=13.5$, $g_A=1.25$, and $f_\pi=0.093$~GeV. The charged-channel isospin factor is $\sqrt2$, and the initial proton spin is averaged. The c.m. differential cross section is
\begin{equation}
\frac{d\sigma}{d\Omega_{\rm cm}}=
 \frac{1}{64\pi^2s}\frac{q_f}{q_i}w_F(s)
 \frac12\sum_{\rm spins}|\mathcal M_0^C|^2.
\label{eq:two_body_cross_section}
\end{equation}
Here $t$ denotes the corresponding invariant amplitude in the spinor convention of Ref.~\cite{Hyodo2012}; the relation is $\mathcal M_0=2\sqrt{M_NM_\Theta}\,t$.

Assuming that $\Gamma_\Theta$ is saturated by the two isospin channels $K^+n$ and $K^0p$ with equal partial widths,
\begin{equation}
g_{KN\Theta}^2=\frac{2\pi M_\Theta\Gamma_\Theta}{q_d(E_N^d-M_N)},\qquad
q_d=\frac{\lambda^{1/2}(M_\Theta^2,M_N^2,m_K^2)}{2M_\Theta},
\label{eq:width_coupling}
\end{equation}
where $E_N^d=\sqrt{M_N^2+q_d^2}$ and $\Gamma_\Theta$ is in GeV. The weight $w_F(s)$ in Eq.~(\ref{eq:two_body_cross_section}) is $w_s$ for the static prescription and $w_c$ for the covariant one; the form-factor weights multiplying the bare squared amplitude are
\begin{align}
F_s&=\frac{\Lambda_s^2}{\Lambda_s^2+q_i^2},&w_s&=F_s^2,&\Lambda_s&=0.500~{\rm GeV},\nonumber\\
F_c&=\frac{\Lambda_c^4}{\Lambda_c^4+(s-M_N^2)^2},&w_c&=F_c^4,&\Lambda_c&=1.800~{\rm GeV}.
\label{eq:formfactors}
\end{align}
Here $q_i=\lambda^{1/2}(s,m_\pi^2,M_N^2)/(2\sqrt{s})$ is the initial c.m. momentum. Both weights depend only on the beam energy and therefore cancel in the normalized angular distribution at fixed beam energy for a given vertex, while PS and PV retain different angular dependences. The positive-parity cross sections, energy dependence, and angular distributions of Ref.~\cite{Hyodo2012} provide a near-threshold consistency check. Small differences from the tabulated values remain; their origin cannot be established uniquely because Ref.~\cite{Hyodo2012} does not specify its pion, nucleon, and kaon mass inputs. Results at energies above those examined in Ref.~\cite{Hyodo2012} are Born-model extrapolations.

\subsection{Negative-parity nucleon-exchange amplitude}

For $J^P=1/2^-$ the same $s$-channel nucleon exchange is used with a scalar $\Theta^+$ vertex. Replacing $\gamma_5$ by the unit matrix at the $KN\Theta$ vertex, and taking the generalized Goldberger-Treiman relation with the opposite mass difference, the amplitudes become
\begin{align}
\mathcal M_0^{\rm PS,-}
 &=\frac{\sqrt2g_{\pi NN}g_{KN\Theta}}{s-M_N^2}
 \bar u_\Theta(\not P_s+M_N)\gamma_5u_p,\nonumber\\
\mathcal M_0^{\rm PV,-}
 &=\frac{\sqrt2g_{KN\Theta}g_A}
 {2f_\pi(M_\Theta-M_N)(s-M_N^2)}
 \bar u_\Theta\not p_K(\not P_s+M_N)\not p_\pi\gamma_5u_p.
\label{eq:born_amplitude_negative}
\end{align}
The pionic vertex, the couplings, and the initial-state spin average are those of Eq.~(\ref{eq:born_amplitude}), and Eqs.~(\ref{eq:two_body_cross_section}) and~(\ref{eq:formfactors}) apply unchanged. The $\Theta^+$ now carries an $s$-wave $KN$ pair, so the coupling is fixed by
\begin{equation}
g_{KN\Theta}^2=\frac{2\pi M_\Theta\Gamma_\Theta}{q_d(E_N^d+M_N)},
\label{eq:width_coupling_negative}
\end{equation}
with $q_d$ and $E_N^d$ as in Eq.~(\ref{eq:width_coupling}). At fixed $\Gamma_\Theta$ this relation gives $g_{KN\Theta}^2$ smaller than Eq.~(\ref{eq:width_coupling}) by the factor $(E_N^d-M_N)/(E_N^d+M_N)\approx1/50$ at the physical decay momentum. The corresponding entries of Table~I of Ref.~\cite{Hyodo2012} are reproduced within the same near-threshold rounding as the positive-parity ones.

\subsection{Laboratory angular transformation}

The c.m. kaon momentum is $q_f=\lambda^{1/2}(s,m_K^2,M_X^2)/(2\sqrt{s})$, where $\lambda(x,y,z)=x^2+y^2+z^2-2xy-2xz-2yz$. Let $\theta^*$ be the c.m. kaon angle relative to the incident pion, $u=\cos\theta^*$, $E_K^*=\sqrt{m_K^2+q_f^2}$, and $a=\beta_{\rm cm}E_K^*$. The laboratory components are
\begin{equation}
p_\parallel=\gamma_{\rm cm}(q_fu+a),\qquad p_\perp=q_f\sqrt{1-u^2},
\label{eq:boost}
\end{equation}
with $\beta_{\rm cm}=p_\pi/(E_\pi+M_N)$ and $\gamma_{\rm cm}=(1-\beta_{\rm cm}^2)^{-1/2}$. The angular distribution transforms as
\begin{equation}
\frac{d\sigma}{d\Omega_{\rm lab}}=
 \sum_b\left|\frac{d\cos\theta_b^*}{d\cos\theta_{\rm lab}}\right|
 \frac{d\sigma}{d\Omega_{\rm cm}}(\theta_b^*),
\label{eq:jacobian}
\end{equation}
where $b$ runs over the physical branches retained by the momentum selection; for $a>q_f$ two branches lie below the maximum laboratory angle, and for $a<q_f$ the map is single valued. The integrals defining $\bar\sigma$ and the accepted yields are evaluated over the c.m. angular intervals that map onto the selected laboratory angular interval on the branches retained by the momentum selection. The former excludes detector efficiencies, while the latter includes the acceptance and efficiency factors of Eqs.~(\ref{eq:signal_counts}) and (\ref{eq:background_counts}).

\end{document}